\documentclass[aps,prb,twocolumn,superscriptaddress]{revtex4-1}
\usepackage{graphicx}
\usepackage[hidelinks]{hyperref} 
\usepackage{bm}
\usepackage{amsmath}

\usepackage{hyperref}
\hypersetup{colorlinks,
}

\graphicspath{{./}}

\begin{document}


\title{Antiferromagnetic long-range order in $5d^1$ double-perovskite Sr$_2$MgReO$_6$}

\renewcommand*{\thefootnote}{\arabic{footnote}}

\author{Shang Gao}
\email[]{shang.gao@riken.jp}
\thanks{Current address: Materials Science \& Technology Division and Neutron Science Division, Oak Ridge National Laboratory, Oak Ridge, TN 37831, USA}
\affiliation{RIKEN Center for Emergent Matter Science, Wako 351-0198, Japan}

\author{Daigorou Hirai}
\email[]{dhirai@issp.u-tokyo.ac.jp}
\affiliation{Institute for Solid State Physics, University of Tokyo, Kashiwa 277-8581, Japan}

\author{Hajime Sagayama}
\affiliation{Institute of Materials Structure Science, High Energy Accelerator Research Organization,
Tsukuba 305-0801, Japan}

\author{Hiroyuki Ohsumi}
\affiliation{RIKEN SPring-8 Center, Sayo 679-5148, Japan}

\author{Zenji Hiroi}
\affiliation{Institute for Solid State Physics, University of Tokyo, Kashiwa 277-8581, Japan}

\author{Taka-hisa Arima}
\affiliation{RIKEN Center for Emergent Matter Science, Wako 351-0198, Japan}
\affiliation{Department of Advanced Materials Science, University of Tokyo, Kashiwa 277-8561, Japan}


\date{\today}


\begin{abstract}
The double-perovskite $A_2BB'$O$_6$ with heavy transition metal ions on the ordered $B'$ sites is an important family of compounds to study the interplay between electron correlation and spin-orbit coupling (SOC). Here we prepared high-quality Sr$_2$MgReO$_6$ powder and single-crystal samples and performed non-resonant and resonant synchrotron x-ray diffraction experiments to investigate its magnetic ground state. By combining the magnetic susceptibility and heat capacity measurements, we conclude that Sr$_2$MgReO$_6$ exhibits a layered antiferromagnetic (AF) order at temperatures below $\sim$ 55~K with a propagation vector $\bm{q} = (001)$, which contrasts the previously suspected spin glass state. Our works clarify the magnetic order in Sr$_2$MgReO$_6$ and demonstrate it as a candidate system to look for magnetic octupolar orders and exotic spin dynamics.
\end{abstract}

\maketitle

\section{Introduction}

The relativistic SOC entangles the spin and angular momenta of electrons and fosters a variety of exotic electronic states. A well-known example is the topological insulator where SOC induces a transfer of the Berry curvature among the electron bands~\cite{bernevig_topo_2013}. For systems with strong electron correlation, SOC plays a more local but equally eminent role~\cite{krempa_correlated_2014}. With an appropriate number of electrons, \textit{e.g.} the Ir$^{4+}$ ions with the $5d^5$ configuration in Sr$_2$IrO$_4$~\cite{kim_novel_2008, kim_phase_2009}, SOC enhances electron correlation and promotes the Mott transition. More importantly, SOC often induces non-Heisenberg superexchange couplings between the local magnetic moments~\cite{kugel_jahn_1982, chen_spin_2008, jackeli_mott_2009}, leading to exotic states like the Kitaev spin liquid with fractional excitations~\cite{jansa_observation_2018} or multipolar orders that preserve the time reversal symmetry~\cite{santini_multipolar_2009, chen_exotic_2010, chen_spin_2011}.

\begin{figure}[t!]
\includegraphics[width=0.45\textwidth]{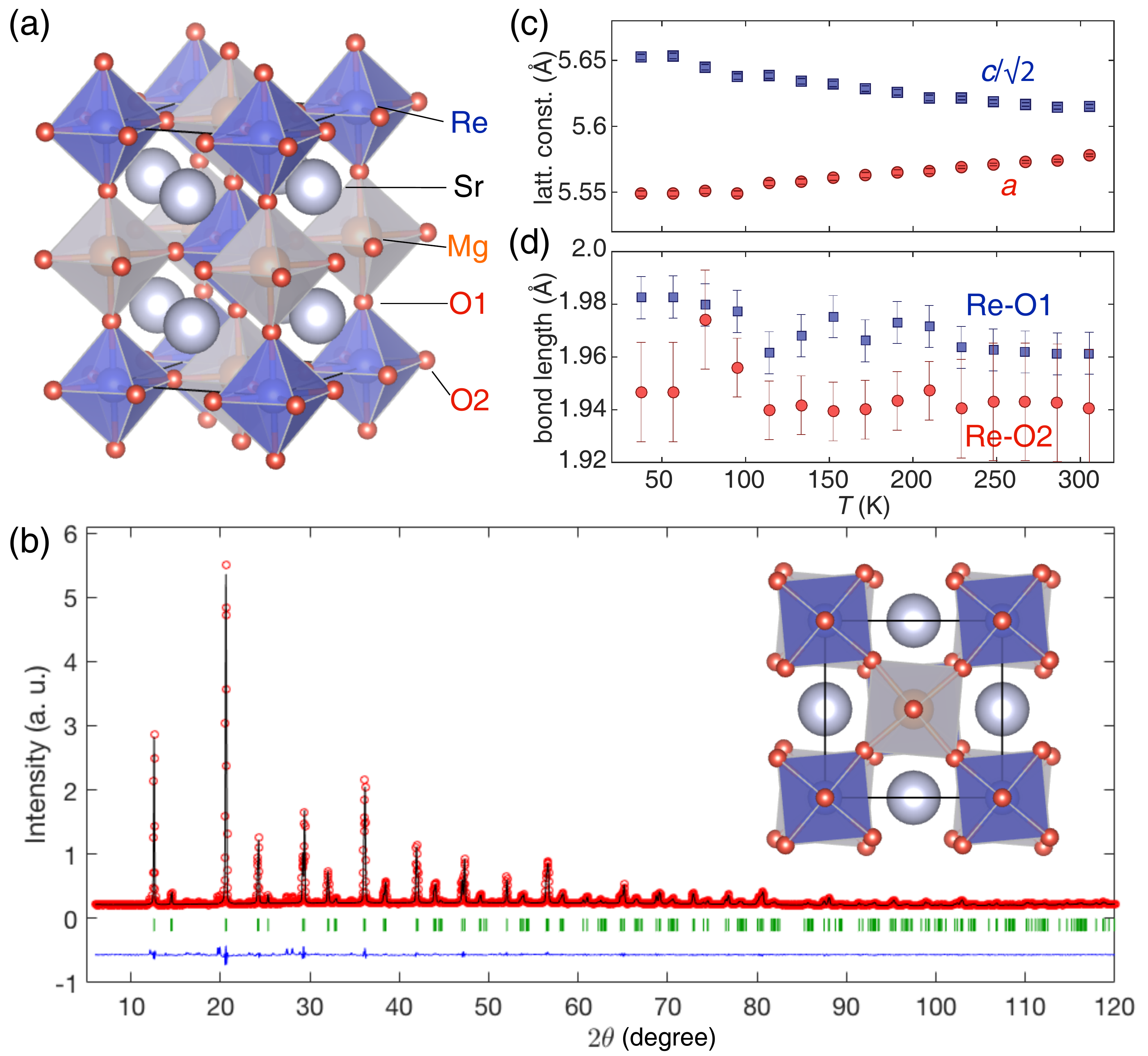}
\caption{(a) Crystal structure of Sr$_2$MgReO$_6$. The MgO$_6$ and ReO$_6$ octahedra are shown explicitly. (b) Refinement result of the synchrotron x-ray diffraction data measured at room temperature for Sr$_2$MgReO$_6$ polycrystalline sample. Data points are shown as red circles. The calculated pattern is shown as the black solid line. The vertical bars show the positions of the Bragg peaks for Sr$_2$MgReO$_6$. The blue line at the bottom shows the difference of measured and calculated intensities. Inset is the Sr$_2$MgReO$_6$ structure viewed along the $c$ axis. (c-d) $T$-dependence of the lattice constants (c) and Re-O bond lengths (d).
\label{fig:powder}}
\end{figure}

The double-perovskite $A_2BB'$O$_6$ (structure shown in Fig.~\ref{fig:powder}(a)) with the $B'$ sites occupied by the heavy transition metal ions is an important family of compounds to study the interplay between electron correlation and SOC.  Its alternately ordered $B$ and $B'$ sites form two interpenetrating face-centered cubic (FCC) lattices, where the reduced overlapping between the $B'$ sites facilitates electron localization. Mean-field analysis for systems with the $d^1$ or $d^2$ electron configuration~\cite{chen_exotic_2010, chen_spin_2011} shows that strong SOC could induce a quadrupolar nematic order and a consequent staggered ferrimagnetic order that might be experimentally realized in Ba$_2$NaOsO$_6$~\cite{erickson_ferromagnetism_2007,lu_magnetism_2017}, Cs$_2$TaCl$_6$~\cite{ishikawa_ordering_2019}, and Ba$_2$MgReO$_6$~\cite{hirai_successive_2019, hirai_detection_2020}. Under dominant AF couplings, the classical long-range ordered (LRO) ground state become unstable due to quantum fluctuations~\cite{chen_exotic_2010}, leading to a spin liquid of random spin-orbit dimers that is possibly realized in Ba$_2$YMoO$_6$~\cite{vries_valence_2010,romhanyi_spin_2017}. Even in the case of the $d^3$ electron configuration, unquenched orbitals and SOC might account for the unusual gapped spin dynamics in Sr$_2$ScOsO$_6$ and Ba$_2$YOsO$_6$~\cite{taylor_spin_2016, taylor_spin_2017}.

In this paper, we focus on the static spin correlations in the double-perovskite Sr$_2$MgReO$_6$, where the Re$^{6+}$ ions are of the $5d^1$ electron configuration. According to the previous studies on powder samples~\cite{wiebe_frustration_2003, kato_structural_2004, greedan_search_2011},  Sr$_2$MgReO$_6$ belongs to the tetragonal space group $I4/m$ with the ReO$_6$ octahedra slightly elongated along the $c$ axis. Below $\sim 50$ K, Sr$_2$MgReO$_6$ is supposed to enter a spin glass state based on the following observations~\cite{wiebe_frustration_2003}: 1.~A broad peak in the magnetic susceptibility at $\sim 50$ K that is accompanied by a weak hump in the heat capacity; 2. Bifurcated magnetic susceptibilities in field cooling (FC) and zero field cooling (ZFC) up to $\sim 300$~K; 3.~Absence of magnetic reflections in powder neutron diffraction measurements. However, the reported magnetic susceptibility exhibits a Curie-Weiss temperature $\Theta_\mathrm{CW}$ of $\sim426$~K that is anomalously high compared to the similar compounds~\cite{chen_exotic_2010}. On the other hand, it is currently known that strong SOC leads to spin and orbital compensation in $5d$ systems and reduces the magnetic moment size, making it difficult to detect magnetic Bragg peaks in conventional powder neutron diffraction experiments~\cite{taylor_magnetic_2015, morrow_spin_2016}.

Here we reinvestigate the ground state of Sr$_2$MgReO$_6$ using high-quality powder and single-crystal samples. At $\sim55$ K, a sudden rise in the magnetic susceptibility together with a jump in the heat capacity are observed, both evidencing a second-order phase transition. No bifurcation exists for the magnetic susceptibilities in FC and ZFC, and the Weiss temperature is fitted to be $\sim 134$~K, which is much lower than the previously reported value. Using resonant x-ray diffraction, we reveal the ground state to be an AF LRO state with a propagation vector $\bm{q} = (001)$, and the moment directions are determined to be within the $ab$ plane.  

\section{Experimental details}\label{sec:disc}

Polycrystalline samples of Sr$_2$MgReO$_6$ were synthesized through conventional solid-state reactions. Stoichiometric amounts of SrO, MgO, and ReO$_3$ were ground together in an argon-filled glove box and then pressed into a pellet. The pellet was wrapped in a gold foil, sealed in an evacuated quartz ampoule, and then heated at 800 $^{\circ}$C for 50 hours. The sintered pellet was reground and pelletized before sintering at 900 $^{\circ}$C for 50 hours. 

Single crystals of Sr$_2$MgReO$_6$ were grown via the molten flux method. SrO, MgO, ReO$_3$, and SrCl$_2$ in a molar ratio of 2:1:1:6 were mixed in a glove box, and the mixture was put in a platinum tube and sealed in an evacuated quartz ampoule. The ampoule was heated to 1400 $^{\circ}$C in 14 hours, held at the temperature for 2 hours, and then slowly cooled to 1000 $^{\circ}$C at a cooling rate of 5 $^{\circ}$C/h. The SrCl$_2$ flux matrix was removed by washing with distilled water. 

Magnetic susceptibility measurements on a sintered pellet of the polycrystalline sample were performed using a superconducting quantum interference device (SQUID, MPMS-III, Quantum Design). Heat capacity was measured using the semi-adiabatic thermal relaxation technique in a physical property measurement system (PPMS, Quantum Design).

Non-resonant synchrotron x-ray diffraction experiments were performed on BL-8A at Photon Factory in Japan. A helium cryogenic gas cooler was employed. For polycrystalline measurements, a glass capillary with a thin layer of Sr$_2$MgReO$_6$ polycrystals on the inner wall was prepared to reduce x-ray absorption.  For the single-crystal measurements, a piece of Sr$_2$MgReO$_6$ single crystal ($~20\times20\times20~\mu m^3$) was selected. The energy of the incident x-rays was tuned to 12.4 keV (15.9 keV) for the polycrystalline (single-crystal) measurements and its exact wavelength was determined as 1.00224(8)~\AA\ (0.77716(6)~\AA) using the CeO$_2$ standard. Refinements for the polycrystalline and single-crystal data were performed with RIETAN~\cite{izumi_three_2007} and SHELXL~\cite{sheldrick_crystal_2015}, respectively.

Resonant x-ray diffraction experiments were performed on BL19LXU at SPring-8 in Japan~\cite{yabashi_design_2001}. The incident x-rays were linearly polarized perpendicular to the scattering plane ($\sigma$), and its energy ($E$) was tuned to 10.535 keV around the Re $L_3$ absorption edge. A piece of Sr$_2$MgReO$_6$ single crystal ($\sim0.8\times0.8\times0.8$~mm$^3$) was attached to a copper sample holder using silver paste and then mounted in a closed cycle $^4$He refrigerator. A PG(008) crystal with a scattering angle of $2\theta = 89.2^{\circ}$ was used to analyze the polarization of the scattered x-rays.

\section{Results}\label{sec:disc}
Throughout the investigated temperature regimes (from $\sim37$~K to 300~K), the synchrotron x-ray diffraction pattern collected on a Sr$_2$MgReO$_6$ polycrystalline sample can be refined using the tetragonal space group $I4/m$. Fig.~\ref{fig:powder}(b) plots a representative refinement result for data collected at 300 K with the $R$-factors $R_p = 1.77$~\% and $R_{wp} = 2.48$~\%. Tiny peaks from the impurity phase, \textit{e.g.}, those at $2\theta \sim 28^{\circ}$, were excluded in the refinement. The refined lattice constants are $a = 5.578(1)$~\AA\ and $c = 7.941$(2)~\AA, and the refined oxygen positions are (0 0 0.247(1)) for O1 and (0.223(2) 0.267(2) 0) for O2. A tiny fraction of $\sim1.1$~\% Re is detected on the Mg sites, while the Re sites are fully occupied.

The temperature ($T$) dependence of the lattice constants extracted from the refinements is shown in Fig.~\ref{fig:powder}(c). Similar to Sr$_2$MgOsO$_6$~\cite{morrow_spin_2016}, the lattice constant expands (shrinks) along the $c$ ($a$) axis with decreasing $T$, which suggests increasing elongation for the ReO$_6$ octahedra. This scenario is confirmed in the $T$-dependence of the Re-O bond lengths shown in Fig.~\ref{fig:powder}(d), where the length of the Re-O1 bonds along the $c$ axis increases with lowering $T$, while the length of the Re-O2 bonds in the $ab$ plane almost stays constant, indicating the shrinkage of the lattice constant $a$ is mainly achieved through the rotation of the ReO$_6$ octahedra around the $c$ axis. 
\begin{figure}[t!]
\includegraphics[width=0.45\textwidth]{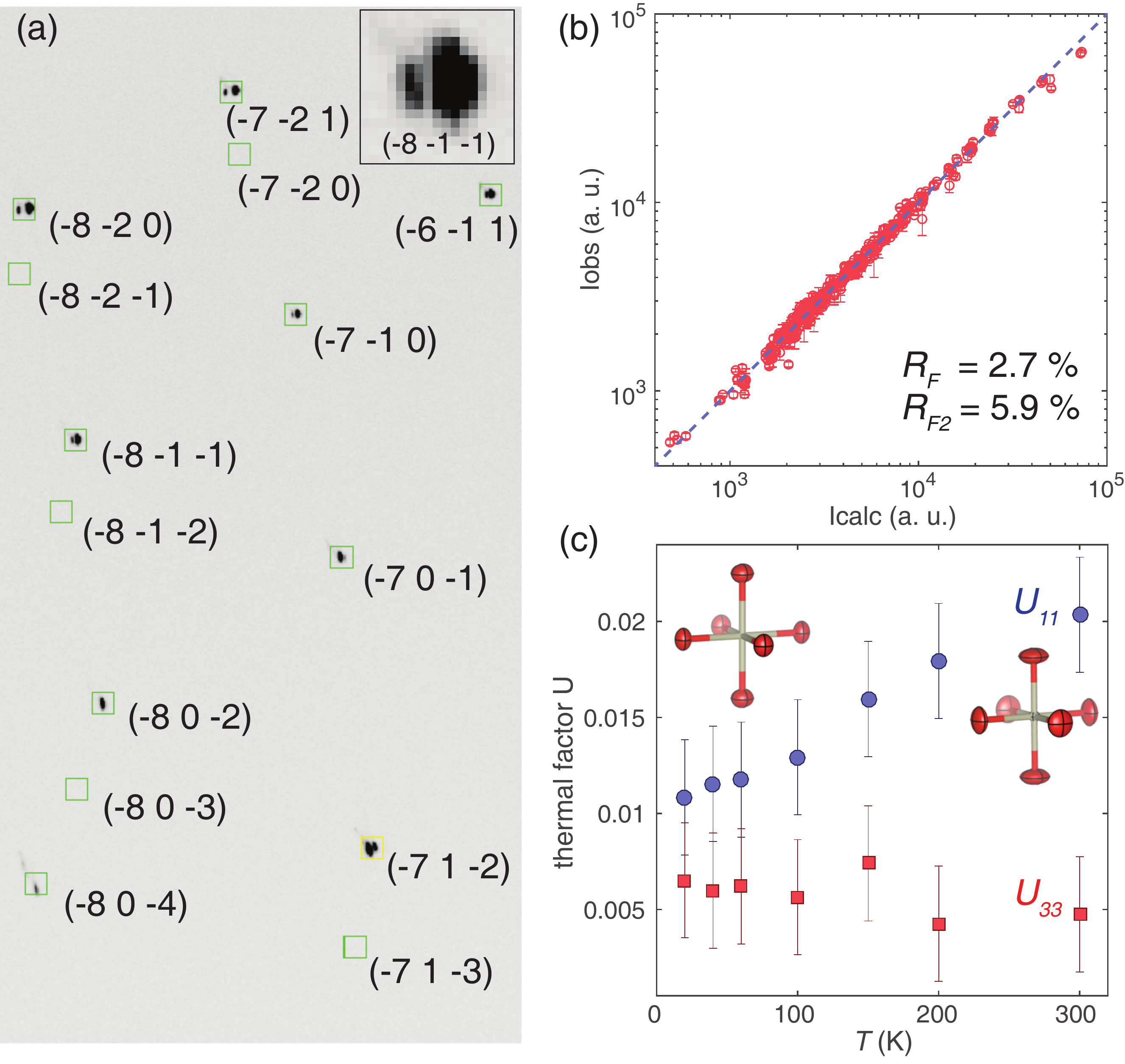}
\caption{(a) X-ray oscillation photograph of Sr$_2$MgReO$_6$ collected at $\sim 38$~K. Inset shows the zoomed-in region around the (-8 -1 -1) reflection. (b) Comparison between the observed intensities at $T=300$~K and the calculated intensities assuming a tetragonal $I4/m$ structure. (c) $T$-dependence of the thermal factors $U_{11}$ and $U_{33}$ for O1. Insets plot the thermal ellipsoids for the six O$^{2-}$ ions around Re$^{6+}$ at 300 K (right) and 38 K (left).
\label{fig:xtal}}
\end{figure}

The tetragonal crystal structure is further confirmed in our non-resonant x-ray diffraction experiment on a single-crystal sample. Fig.~\ref{fig:xtal}(a) is a representative x-ray oscillation photograph of Sr$_2$MgReO$_6$ collected at $\sim 38$~K, where the peaks split due to the tetragonal crystallographic twins. Compared to the data collected at higher temperatures, no additional superlattice reflections were observed at $\sim 38$~K, indicating that the transition at $\sim55$~K does not involve structural distortions. Refinements of the single-crystal x-ray diffraction dataset were performed in the tetragonal $I4/m$ space group by summing up the intensities from different twins. Fig.~\ref{fig:xtal}(b) compares the integrated intensities measured at 300 K with the calculated intensities. The $R$-factors of $R_F = 2.7\ \%$ and $R_{F2} = 5.9\ \%$ confirm the high quality of our crystals. The refined oxygen positions for O1: (0 0 0.241(2)) and O2: (0.218(2) 0.264(2) 0) are similar to the refinement results for the polycrystalline sample. As shown in Fig.~\ref{fig:xtal}(c), the thermal factors for the O$^{2-}$ ions are highly anisotropic at 300~K, which indicates the rotation of the rigid ReO$_6$ octahedra as the main contribution to thermal fluctuations at high temperatures.

\begin{figure}[t!]
\includegraphics[width=0.475\textwidth]{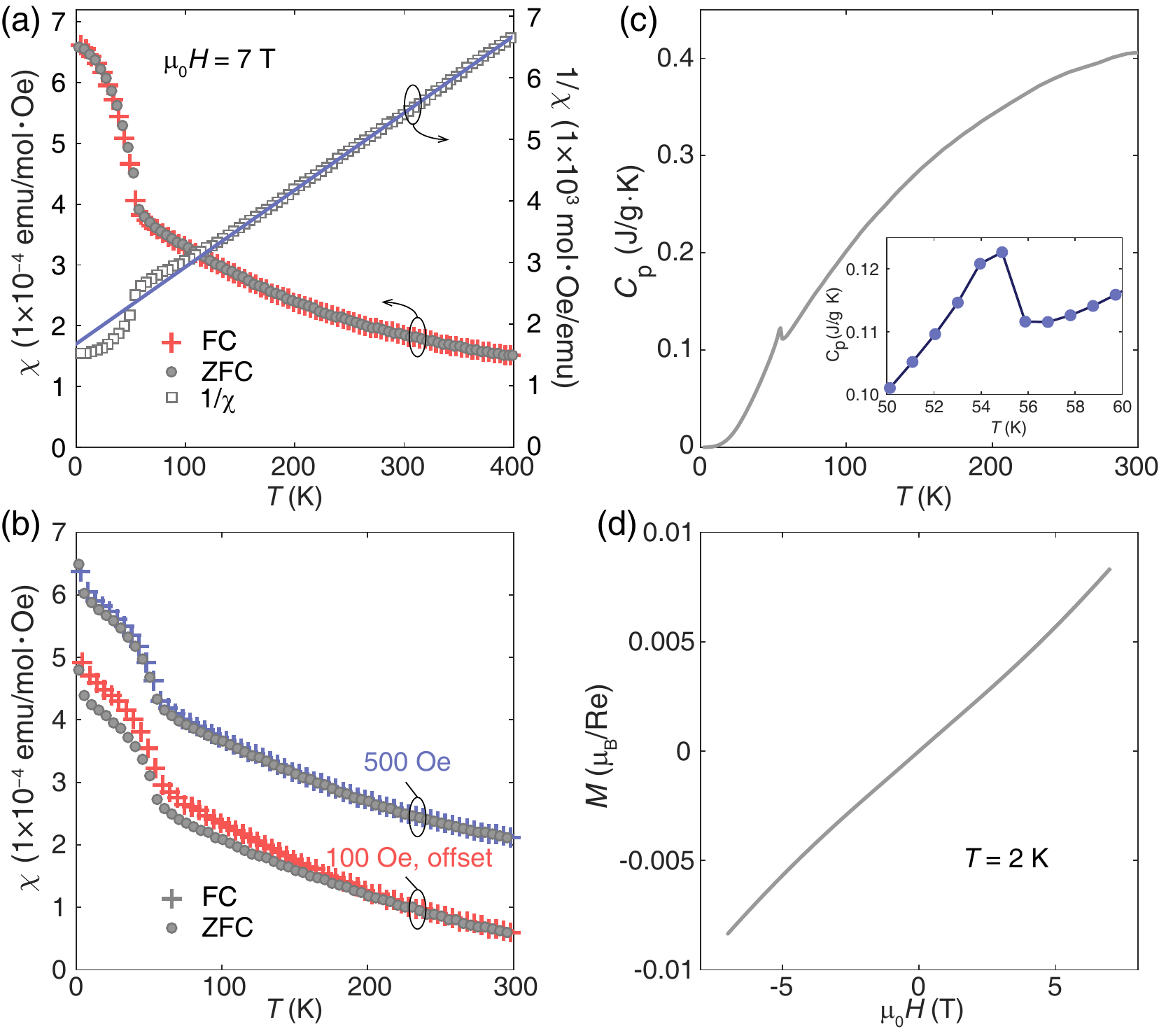}
\caption{(a) Left axis shows the $T$-dependence of the magnetic susceptibility $\chi(T)$ measured in zero field cooling (ZFC, filled circles) and 7-T field cooling (FC, crosses). The applied field is 7 T. Right axis shows the $T$-dependence of the reversed magnetic susceptibility measured in ZFC (empty squares). Blue solid line is fit to the Curie Weiss law. Measurements were performed on a polycrystalline sample. (b) $T$-dependence of the heat capacity $C_p$. Inset plots the zoomed-in region around 55 K. (c) $\chi(T)$ measured in a low field of 100 and 500 Oe. The 100 Oe data is offset by $-1.5\times10^{-4}$ emu/mol$\cdot$Oe for clarity. (d) Field-dependence of the magnetization $M(H)$ measured at $T = 2$~K.
\label{fig:transport}}
\end{figure}

The $T$-dependence of the magnetic susceptibility $\chi(T)$ measured in a 7-T field using a polycrystalline sample is shown in Fig.~\ref{fig:transport}(a). From 400 K down to the base $T$ of 2 K, there is no obvious deviation between the FC and ZFC data. As shown in Fig.~\ref{fig:transport}(b), the deviation only becomes apparent in a low measuring field of 100 Oe. Such a deviation might arise from the weak contributions of a tiny amount of impurity phase. With lowering $T$, the magnetic susceptibility increases sharply at  $T_N \approx 55$~K, indicating the appearance of a magnetic transition. Such a transition is also evident as a jump in the heat capacity presented in Fig.~\ref{fig:transport}(c) that is characteristic of a second-order phase transition. The entropy change across the phase transition can be estimated to be 1.57~J/Kmol by assuming a linear background between 40 and 60 K. This value is only $\sim27$~\% of $R\ln2$ and indicates the existence of strong short-range correlations above $T_N$. As shown in Fig.~\ref{fig:transport}(a), the inverse magnetic susceptibility obeys the Curie-Weiss law down to $\sim120$~K. The fitted effective moment is $\mu_{\rm{eff}} = 0.8\ \mu_B$, which is compatible with a reduced moment size due to SOC. The fitted Weiss temperature of $\Theta_{CW} = -134$~K is close to the temperature at which $\chi(T)$ starts to deviate from the Curie Weiss law. The ratio of $\Theta_{CW}/T_N\approx 2.4$ is consistent with the weak frustration character of the FCC lattice. 

Although the magnetic susceptibility increases below $T_N$, the field-dependence of the magnetization $M(H)$ shown in Fig.~\ref{fig:transport}(d) does not exhibit hysteresis and its slight nonlinear behavior can be ascribed to the magnetic anisotropy. The absence of hysteresis indicates the AF character of the ground state, which contrasts the ferromagnetic (FM) components observed in Ba$_2$MgReO$_6$~\cite{hirai_successive_2019} and Ba$_2$NaOsO$_6$~\cite{erickson_ferromagnetism_2007}, where a canted ferromagnetic structure with the propagation vector $\bm{q} = (001)$ has been proposed. In this structure, magnetic moments are parallel within the $ab$ layers, but canted between the neighbouring $ab$ planes. Therefore, for Sr$_2$MgReO$_6$, a candidate magnetic structure might be a similar layered AF structure without canting. The slightly increased magnetic susceptibility in the LRO phase might be due to the parasitic ferromagnetism caused by the surfaces of the grains or the imperfect ordering of the Mg and Re ions, or even the ferromagnetic domain walls that have been observed in other $5d$ compounds~\cite{disseler_magnetic_2012, tardif_all_2015, ma_mobile_2015}. Assuming the intrinsic magnetic susceptibility of Sr$_2$MgReO$_6$ at $T = 0$~K to be $2/3$ of the value at $T_N$ as expected for the conventional antiferromagnets, the contributions of the parasite ferromagnetism  at 0~K can be estimated to be $\sim4\times10^{-4}$~emu/mol$\cdot$Oe. 

\begin{figure}[t!]
\includegraphics[width=0.45\textwidth]{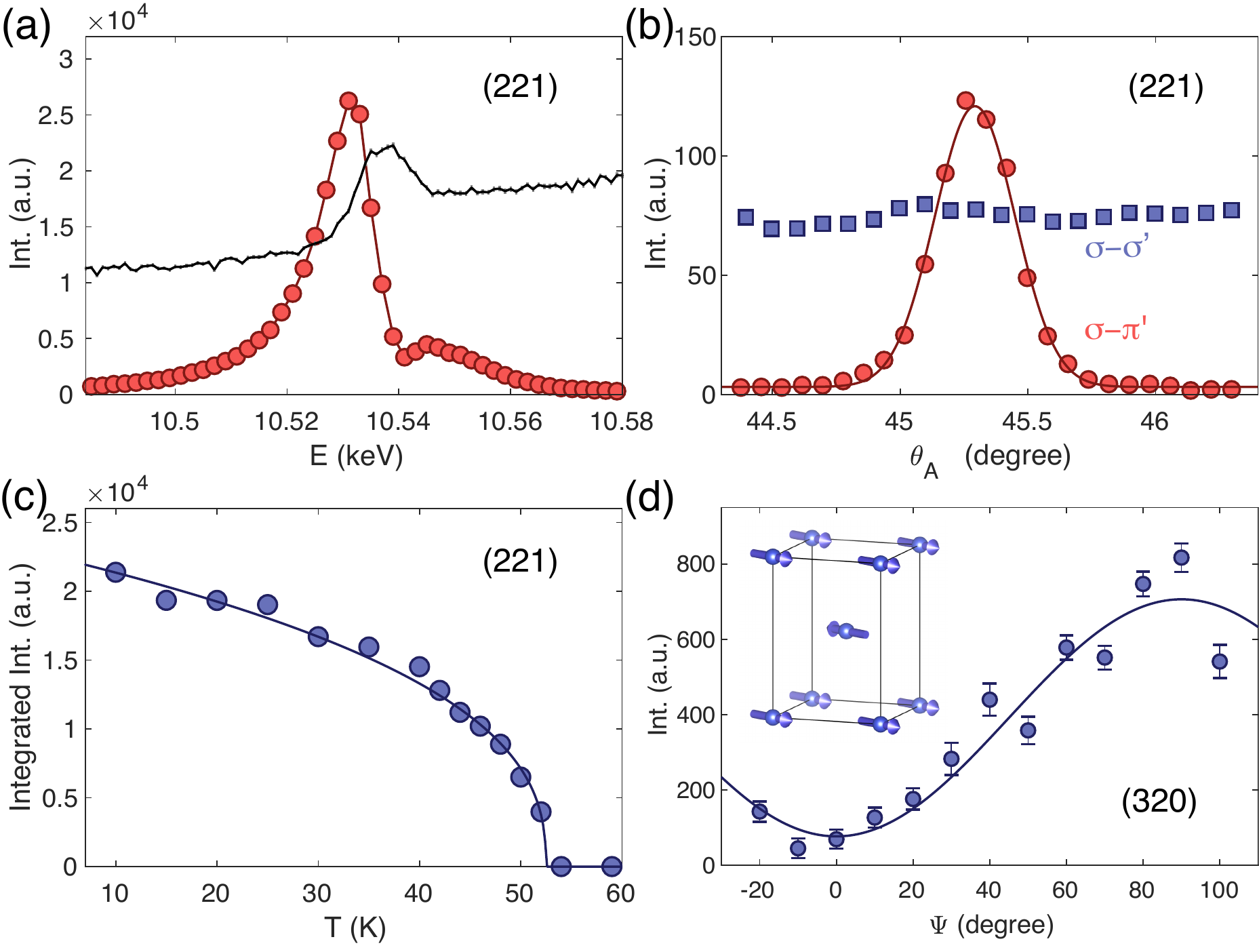}
\caption{(a) Energy dependence of the intensity for the (221) reflection around the Re $L_3$ edge measured at $\sim 6$ K (red circles). The black solid line is the x-ray absorption spectrum. (b) Polarization analysis of the scattered intensities around (221). $\theta_{A}$ is the angle of the analyzer. (c) $T$-dependence of the intensity for the (221) reflection. The solid line is a fit to a power law, where the integrated intensities are proportional to $(T_N-T)^\eta$ with $T_N = 52.5(2)$ and $\eta = 0.38(2)$. (d) Azimuthal angle $\Psi$-dependence of the intensity for the (320) reflection measured at $\sim 6$~K. Solid line is the fit to the $A\sin^2\Psi+C$ relation. Inset plots a representative magnetic structure with the Re$^{6+}$ moments in the $ab$ plane.
\label{fig:rxd}}
\end{figure}

In order to clarify the magnetic order, we resorted to the resonant x-ray diffraction technique. In proximity to the absorption edges, the energy dependence of the x-ray scattering cross section is strongly affected by the local charge and spin density distributions, lending an opportunity to detect the delicate ordering of charges, spins, and orbitals through resonant x-ray diffraction~\cite{murakami_resonant_2017}. Specifically, for dipole-dipole  (E1-E1) transitions, the resonant x-ray scattering cross section can be written as
\begin{equation}
\begin{aligned}
f_{E 1}^{\mathrm{res}} &\propto \boldsymbol{\epsilon}^{\prime} \cdot \boldsymbol{\epsilon}\left[F_{11}+F_{1-1}\right] -i \boldsymbol{\epsilon}^{\prime} \times \boldsymbol{\epsilon} \cdot \bm{\hat{z}}\left[F_{11}-F_{1-1}\right] \\
&+\left(\boldsymbol{\epsilon}^{\prime} \cdot \bm{\hat{z}}\right)(\boldsymbol{\epsilon} \cdot \bm{\hat{z}})\left[2 F_{10}-F_{11}-F_{1-1}\right]\ \rm{,}
\end{aligned}
\label{eqn:cross}
\end{equation}
where $\bm{\epsilon}$ ($\bm{\epsilon}^{\prime}$) is the polarization of the incident (scattered) x-rays, $\bm{\hat{z}}$ is the unit vector along the magnetic moment, $F_{1j}$ ($j = \pm1$ or 0) corresponds to the E1 transition that induces a change of $j$ in the $\bm{\hat{z}}$ component of the total orbital angular momentum. Therefore, for incoming x-rays fixed to the $\sigma$ polarization, magnetic scattering will appear only in the $\sigma$-$\pi'$ channel but not in the $\sigma$-$\sigma'$ channel due to the $\bm{\epsilon}'\times \bm{\epsilon}$ coefficient in Eqn.~(\ref{eqn:cross}).

With an incident x-ray energy of 10.535~keV that is close to the Re $L_3$ absorption edge, we performed scans in reciprocal space to search for magnetic reflections. A series of reflections were observed at positions that are forbidden in the $I4/m$ space group, including (010), (032), (120), (122), (131), (133), (221), (223), and (232), which suggests a magnetic propagation vector $\bm{q}=(001)$ and thus a layered AF structure. Fig.~\ref{fig:rxd}(a) plots the energy dependence for a representative reflection (221) together with the x-ray absorption spectrum. The intensity of the (221) reflection is strongly enhanced at energies close to the Re $L_3$ absorption edge, indicating its magnetic origin. Such a magnetic origin is further corroborated by the polarization analysis shown in Fig.~\ref{fig:rxd}(b), where the reflection is only observable in the $\sigma$-$\pi'$ channel as expected for magnetic reflections. As shown in Fig.~\ref{fig:rxd}(c), the (221) reflection disappears at temperatures above $\sim 52$~K, which is consistent with the transition temperature of 55~K revealed in $\chi(T)$ and $C_p(T)$.

The moment directions of the AF structure in Sr$_2$MgReO$_6$ can be constrained through the azimuthal angle $\Psi$ scans presented in Fig.~\ref{fig:rxd}(d), where $\Psi$ is defined to be zero when the (001) direction is within the scattering plane. According to Eqn.~(\ref{eqn:cross}), for $\sigma$ polarized incident x-rays, the scattering intensity is the highest (lowest) when the magnetic moment is parallel with (perpendicular to) the propagation vector of the scattered x-rays. As shown in Fig.~\ref{fig:rxd}(d), the intensity of the (320) reflection is the strongest when the (001) direction is perpendicular to the propagation vector of the scattered x-rays and almost becomes zero when the (001) direction is lying within the scattering plane, leading to a $\sin^2\Psi$-like dependence as indicated by the solid line. Therefore, the moment directions should be perpendicular to the (001) direction, although their exact directions within the $ab$ plane cannot be uniquely determined due to the possible presence of $90^\circ$ magnetic domains in a tetragonal system. A schematic for the Sr$_2$MgReO$_6$ magnetic structure is presented in the inset of Fig.~\ref{fig:rxd}(d).

\begin{figure}[t!]
\includegraphics[width=0.45\textwidth]{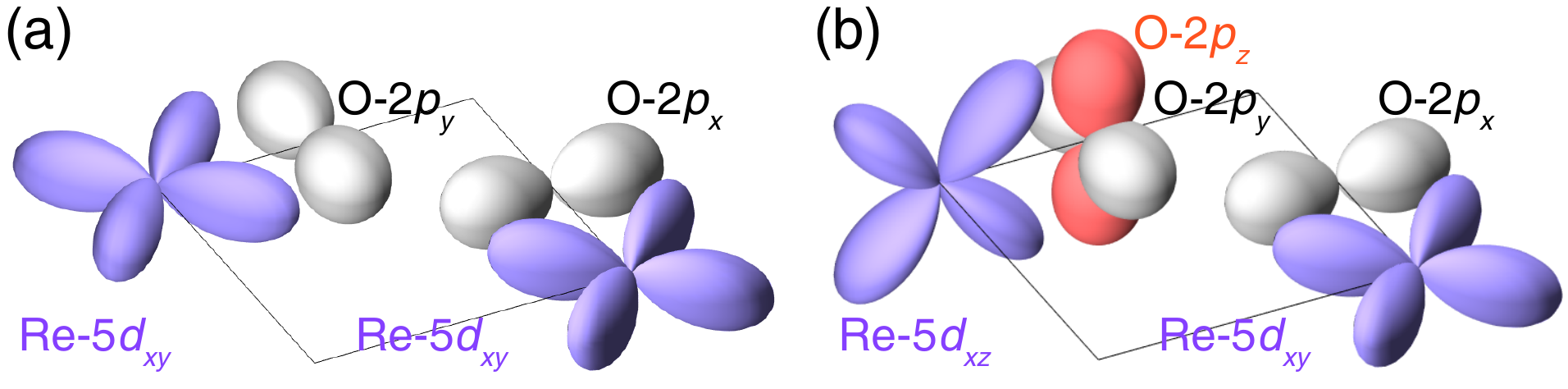}
\caption{Re-O-O-Re exchange paths for the AF (a) and FM (b) couplings between the nearest-neighbouring Re$^{6+}$ ions. This figure is adapted from Ref.~\cite{chen_exotic_2010}. 
\label{fig:path}}
\end{figure}

\section{Discussions}
The ground state of the $5d^1$ double-perovskites in cubic symmetry has been thoroughly investigated in a parameter space that contains three basic nearest-neighbouring (NN) interactions~\cite{chen_exotic_2010}: the AF coupling $J$ ($>0$), the FM coupling $J'$ ($<0$), and the quadrupolar coupling $V$. As shown in Fig.~\ref{fig:path}, the different characters of the NN couplings arise from the exchange paths that involve different orbitals~\cite{chen_exotic_2010}. For the coupling $J$, only one $p$ orbital is involved for each of the O$^{2-}$ ions along the Re-O-O-Re exchange path, leading to its AF character; While for the coupling $J'$, two $p$-orbitals are involved at one of the O$^{2-}$ ions, giving rise to its FM character due to the Hund coupling. Under a dominant $J$, a mean-field calculation predicts a layered AF ground state similar to what we observed in Sr$_2$MgReO$_6$. According to the calculation~\cite{chen_exotic_2010}, the transition temperature $T_N$ and the Curie-Weiss temperature $\Theta_\mathrm{CW}$ can be expressed as
\begin{equation}
\begin{aligned}
&T_N = \frac{J-10J'+\sqrt{73J^2+164J|J'|+100(J')^2}}{36}\ \mathrm{,} \\
&\Theta_{CW} = -\frac{J}{5}- \frac{32J'}{45}\ \mathrm{.}
\end{aligned}
\label{eqn:coupling}
\end{equation}
From our measured $T_N$ and $\Theta_\mathrm{CW}$, the couplings are estimated as $J = 34.2$~meV and $J' = 7.5$~meV if Eqn.~(\ref{eqn:coupling}) is applied. The antiferromagnetic $J'$ suggests that the expressions of $T_N$ and $\Theta_{CW}$ should be modified in Sr$_2$MgReO$_6$. This inadequacy might be due to the tetragonal lattice distortion or the rotation of the ReO$_6$ octahedra around the $c$ axis~\cite{ishizuka_magnetism_2014}, which could induce anisotropic couplings  and thus affect both $T_N$ and $\Theta_\textrm{CW}$.

In the extreme case of $J' = V = 0$, theoretical calculations have revealed that the AF order obtained under the mean-field approximation might be destabilized by quantum fluctuations~\cite{chen_exotic_2010, romhanyi_spin_2017}. Therefore, when quantum fluctuations are properly considered, there should be a phase boundary that separates the AF ordered state from the quantum disordered state centered around the $J' = V = 0$ point. Meanwhile, the absence of magnetic long-range order in the previous study~\cite{wiebe_frustration_2003} suggests that the ground state of Sr$_2$MgReO$_6$ is sensitive to perturbations from disorder, as is often observed in systems lying close to phase boundaries~\cite{chen_magnetic_2016}. Therefore, Sr$_2$MgReO$_6$ might be in proximity to the quantum disordered phase in spite of the observed AF ordered ground state.

The AF order observed in Sr$_2$MgReO$_6$ might be highly non-trivial. SOC entangles the spin and orbital momenta, and induces effective interactions that are of fourth or sixth order in spin operators~\cite{chen_exotic_2010}. Therefore, quadrupolar or octupolar order might develop or even compete with the magnetic dipolar order. SOC and the consequent multipolar interactions also strongly affects the spin dynamics. Different from the conventional two-sublattice antiferromagnets with only two overlapping magnon modes, calculations based on the SU(4) spin wave theory have revealed six different dispersive modes~\cite{chen_exotic_2010}. Observation of these additional modes through experimental techniques like inelastic neutron scattering will provide direct evidence for the entangled spin-orbital excitations in Sr$_2$MgReO$_6$ and further clarify the effect of SOC.

\section{Conclusions}

Through combined magnetic susceptibility, heat capacity, and synchrotron x-ray diffraction measurements, we investigated the static spin correlations of the $5d^1$ double-perovskite compound Sr$_2$MgReO$_6$. Contrary to the previously proposed spin glass state, we observed a magnetic long-range order transition at $\sim55$ K, and a layered two-sublattice AF structure with $\bm{q} = (001)$ was established. The observed AF structure indicates dominant AF couplings in Sr$_2$MgReO$_6$, and suggests it as a candidate compound to look for exotic multipolar orders and entangled spin-orbital excitations.

\begin{acknowledgments}
We thank G. Chen for helpful discussions. We acknowledge T. Nakajima and D. Hashizume for help in the x-ray diffraction experiments. The resonant x-ray diffraction experiment was performed on the BL19LXU beamline at SPring-8 under RIKEN proposals No. 20180095 and 20190073. The off-resonant x-ray diffraction experiments were performed on BL-8A at Photon Factory of High Energy Accelerator Research Organization (KEK) under proposal No. 2019G145. This work was partly supported by Japan Society for the Promotion of Science (JSPS) KAKENHI Grant Number JP18K13491, JP18H04308 (J-Physics), JP19H05826, JP20H01858, and by Core-to-Core Program (A) Advanced Research Networks.
\end{acknowledgments}

\end{document}